\documentclass[conference]{IEEEtran}
\IEEEoverridecommandlockouts
\usepackage{cite}
\usepackage{amsmath,amssymb,amsfonts}
\usepackage{graphicx}
\usepackage{algorithm}
\usepackage{algpseudocode}
\usepackage{textcomp}
\usepackage{url}
\usepackage{multirow}
\usepackage{tcolorbox}
\usepackage{booktabs}
\usepackage{tcolorbox}
\tcbuselibrary{skins}
\usepackage{multirow}
\usepackage{bbding}
\usepackage{utfsym}
\usepackage{ulem}
\usepackage{balance}
\usepackage{wasysym}
\usepackage{hyperref}
\usepackage{multirow}
\usepackage{makecell}
\usepackage{pifont}

\usepackage[utf8]{inputenc}
\usepackage[T1]{fontenc}

\def\eg{\textit{e.g.,} }
\def\ie{\textit{i.e.,} }

\newtcolorbox{AcademicBox}[1][]{academicbox=#1}
\definecolor{SoftBlue}{RGB}{135, 206, 250}  
\definecolor{SoftOrange}{RGB}{255, 224, 178} 
\definecolor{SoftGreen}{RGB}{144, 238, 144}  
\definecolor{CorrectGreen}{RGB}{76, 175, 80} 
\definecolor{ErrorRed}{RGB}{211, 47, 47} 

\def\BibTeX{{\rm B\kern-.05em{\sc i\kern-.025em b}\kern-.08em
    T\kern-.1667em\lower.7ex\hbox{E}\kern-.125emX}}

\tcbset{
  my box/.style={
    enhanced,
    colframe=#1!80,
    colback=#1!10,
    attach boxed title to top left={xshift=0.2cm, yshift=-0.2cm},
    boxed title style={
      colback=#1!80,
      outer arc=0pt,
      arc=0pt,
      top=0pt,
      bottom=0pt,
    },
  },
}

\newtcolorbox{Prompt}[1]{
  my box=black,
  title=#1,
  boxrule=1.2pt,top=6pt,bottom=3.5pt,left=6pt,right=6pt
}

\begin{document}

\def\eg{\textit{e.g.,} }
\def\ie{\textit{i.e.,} }

\newcommand{\td}[1]{{\color{blue}{\textbf{[#1]}}}}


\title{Automatic Multi-level Feature Tree Construction for Domain-Specific Reusable Artifacts Management}

%


 \author{
 \IEEEauthorblockN{Dongming Jin$^{1,2}$, Zhi Jin*$^{1,2}$ \thanks{*Corresponding authors}, Nianyu Li$^{3}$, Kai Yang$^{3}$, Linyu Li$^{1,2}$, Suijing Guan$^{4}$}
    \IEEEauthorblockA{$^1$ School of Computer Science, Peking University, China}
    \IEEEauthorblockA{$^2$ Key Lab of High-Confidence of Software Technologies (PKU), Ministry of Education, China}
    \IEEEauthorblockA{$^3$ ZGC National Laboratory, China}
    \IEEEauthorblockA{$^4$ School of Information Science and Technology, Beijing Forestry University, China}
    \IEEEauthorblockA{\texttt{correspondence to: zhijin@pku.edu.cn}}
}


\maketitle

\begin{abstract}
With the rapid growth of open-source ecosystems (\eg Linux) and domain-specific software projects (\eg aerospace), efficient management of reusable artifacts is becoming increasingly crucial for software reuse. The multi-level feature tree enables semantic management based on functionality and supports requirements-driven artifact selection. However, constructing such a tree heavily relies on domain expertise, which is time-consuming and labor-intensive. 

To address this issue, this paper proposes an automatic multi-level feature tree construction framework named {\sc FTBuilder}, which consists of three stages. \ding{182} It automatically crawls domain-specific software repositories and merges their metadata to construct a structured artifact library. \ding{183} It employs clustering algorithms to identify a set of artifacts with common features. \ding{184} It constructs a prompt and uses LLMs to summarize their common features. {\sc FTBuilder} recursively applies the identification and summarization stages to construct a multi-level feature tree from the bottom up. To validate {\sc FTBuilder}, we conduct experiments from multiple aspects (\eg tree quality and time cost) using the Linux distribution ecosystem. Specifically, we first simultaneously develop and evaluate 24 alternative solutions in the {\sc FTBuilder}. Then we construct a three-level feature tree using the best solution among them. Compared to the official feature tree, our tree exhibits higher quality, with a 9\% improvement in the silhouette coefficient and an 11\% increase in GValue. Furthermore, it can save developers more time in selecting artifacts by 26\% and improve the accuracy of artifact recommendations with GPT-4 by 235\%. {\sc FTBuilder} can be extended to other open-source software communities and domain-specific industrial enterprises.~\footnote{Our code: \url{https://github.com/jdm4pku/FTBuilder}}

\end{abstract}

\section{Introduction}

Software reuse has become a widely adopted practice in modern software development~\cite{frakes2005software}~\cite{lim1994effects}. It aims to utilize existing software artifacts (\eg code snippets and software packages) to build new software, which can reduce development costs and enhance productivity~\cite{mohagheghi2004empirical}~\cite{mohagheghi2007quality}. With the rapid growth of the open-source software ecosystem and the accumulation of domain-specific software artifacts, the number of reusable artifacts has increased exponentially. For example, 10,518,566 new packages were published on the JavaScript node package manager ecosystem in 2023~\cite{wittern2016look}~\cite{web:npm}. In addition, as the scale and complexity of software projects continue to grow, the number of reused artifacts in a project has also increased sharply~\cite{mockus2000case}. For instance, CentOS version 7 contains 14,479 software packages, while version 3.7 released in 2006 contains only 1,275 packages~\cite{jin2024first}. These two factors present a challenge: how to efficiently organize a large number of domain-specific reusable software artifacts and allow developers to locate artifacts based on their requirements quickly.

\begin{figure}
    \centering
    \includegraphics[width=0.7\linewidth]{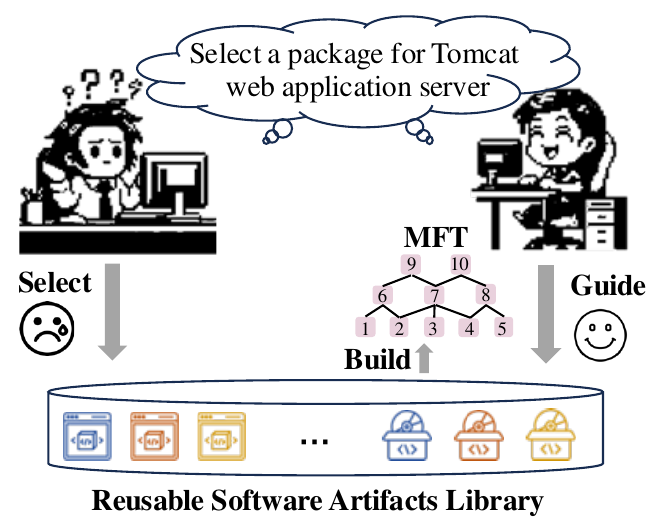}
    \caption{An example of selecting reusable software artifacts by developers.}
    \label{fig:teaser}
\end{figure}

As shown in Figure~\ref{fig:teaser}, it is difficult for developers to directly select artifacts that can satisfy their requirements from a reusable software artifact library. This is because the library contains a large number of artifacts, and there is a mismatch between the granularity in human requirements and the functional descriptions of artifacts~\cite{bakar2015feature}. The multi-level feature tree~\cite{reiser2007multi} can be an effective solution for managing these artifacts. It organizes them into a tree structure based on their functionalities, which can capture both high-level and low-level details about an artifact. Thus, it can provide clear navigation for developers to select artifacts.

However, constructing feature trees requires significant time and effort from domain experts to identify and summarize features. Many studies~\cite{hariri2013supporting}~\cite{guzman2014users} have explored automated requirements feature extraction. On the one hand, these works~\cite{ferrari2013mining}~\cite{weston2009framework} focus on extracting features from requirement documents for forward reuse but give limited attention to reverse extraction from reusable artifact descriptions~\cite{davril2013feature}. On the other hand, existing feature extraction methods primarily rely on traditional natural language processing techniques, such as algebraic models~\cite{guzman2014users}~\cite{kumaki2012supporting}, text preprocessing~\cite{hariri2013supporting}~\cite{yu2013mining}, and term weighting~\cite{acher2012extracting}. These methods lack sufficient performance and practical support tools.

Recently, large language models (LLMs) such as ChatGPT and DeepSeek have exhibited remarkable abilities in natural language understanding and text summarization~\cite{kojima2022large}. LLMs have been used for various requirements-related tasks including requirements elicitation~\cite{gorer2023generating}, modeling~\cite{jin2024evaluation}~\cite{camara2023assessment}, validation~\cite{lubos2024leveraging}, and specification~\cite{lutze2024generating}. They have also demonstrated remarkable performance in these requirements-related tasks. 

Thus, this paper proposes an automatic multi-level feature tree construction framework named {\sc FTBuilder} based on LLMs. The {\sc FTBuilder} consists of three stages. \textbf{1. Artifact Library Construction.} The artifacts' metadata (\eg names and functional descriptions) is crawled from open-source ecosystems or domain-specific software projects. Since software projects may be developed by different teams, the crawled artifacts with the same functionality may have different metadata. Thus, {\sc FTBuilders} employs multiple LLMs to examine the crawled metadata and merge these cases. This stage results in a standardized artifact library. \textbf{2. Common Feature Identification.} {\sc FTBuilder} utilizes an embedding model (\eg \textit{all-MiniLM}) to convert artifacts' functional descriptions into embedding vectors as semantic representations. The vectors are passed into a clustering algorithm (\eg \textit{k-means}) to identify a set of artifacts that contain common functional features. \textbf{3. Common Feature Summarization.} The functional descriptions of the artifacts containing common features are fed into an LLM, and a prompt is constructed to allow the LLM to summarize high-level common features (\ie names and descriptions). {\sc FTBuilder} recursively applies the identification and summarization stages and builds a multi-level feature tree from the bottom up. 

We conduct experiments from multiple aspects to evaluate our {\sc FTBuilder} using a Linux software package ecosystem. \textbf{(1) Empirical Study.} Given that multiple advanced choices (\eg embedding models and clustering algorithms) can be adopted in {\sc FTBuilder}, we simultaneously develop and evaluate 24 alternative solutions. We employ two evaluation metrics (\ie silhouette coefficient~\cite{vrezankova2018different} and Gvalue score~\cite{jin2024first}). Results demonstrate that the best solution is to use
text-embedding-002 to obtain embedding vectors, GMM for
clustering, BIC to select the number of clusters, and GPT-4 to summarize features (Table \ref{tab:empirical}). \textbf{(2) Tree Quality.} The best solution constructs a three-level feature tree
with 201 nodes (\ie features). We compare it with the official Linux feature tree~\cite{web:rpmfinder}. Results show that it outperforms the official tree by 9\% in silhouette coefficient and 11\% in GValue. \textbf{(3) Artifact Selection Time.}  We create a dataset named {\sc ArtSel} to simulate artifact selection based on requirements. We evaluate the compare cost time of three developers on selecting the correct artifact using the two trees. Results show that our constructed tree reduces the average time by 26\%. \textbf{(4) Artifact Recommendation Accuracy.} We compare the accuracy of GPT-4 in artifact recommendation with the guidance of the two trees on the {\sc ArtSel}. We find the constructed feature tree can improve the accuracy by 235\%. 

\textbf{Future research plans.} This current evaluation is only within an open-source ecosystem and an LLM (\ie GPT-4). Future research will extend this evaluation to other ecosystems (\eg JavaScript) and LLMs (\eg DeepSeek).

We summarize our contributions in this paper as follows. 

\begin{itemize}
    \item We propose an automatic multi-level feature tree construction framework named {\sc FTBuilder} based on LLMs. 
    \item We develop 24 alternative solutions under the {\sc FTBuilder} framework and make them available.
    \item We construct an artifact reuse dataset named {\sc ArtSel} that consists of 15 real requirement-artifacts pairs.
    \item We conduct experiments from multiple aspects using a Linux distribution ecosystem. Results show the effectiveness of our {\sc FTBuilder}.
\end{itemize}

\textbf{Data Availability}. We open-source our replication package~\cite{web:code}, which includes the source code and constructed trees. We hope to enable other researchers and practitioners to replicate our work and use it in projects they care about.

\section{background and related works}

\subsection{Software Reuse and Artifact Management}
Software reuse is a key method for enhancing software development efficiency and quality~\cite{mili1995reusing}~\cite{gill2006importance}. The core idea is to build new systems by reusing existing software artifacts or design patterns. Therefore, it can reduce redundant development and improve system reliability~\cite{mohagheghi2004empirical}~\cite{mohagheghi2007quality}. The foundation of software reuse is modular design, which divides software systems into replaceable and reusable artifacts. These artifacts can be code snippets, classes, software packages, and subsystems~\cite{isakowitz1996supporting}. There have been several studies to explore software reusability in practice~\cite{tomer2004evaluating}~\cite{rothenberger2003strategies}~\cite{lee1997empirical}. With the rapid growth of open-source software ecosystems, the efficient management of reusable artifacts has become a critical challenge~\cite{kazman1996scenario}. Traditional methods typically store and retrieve reusable artifacts through artifact libraries or code repositories, but these methods suffer from low retrieval efficiency. This is because artifacts are stored in a flat structure and lack a multi-level requirements feature management, which makes it difficult for developers to locate appropriate artifacts quickly. To address this issue, this paper aims to manage reusable artifacts through the automated construction of multi-level feature trees, improving the efficiency of managing large-scale reusable artifacts.

\subsection{Feature Tree Construction}
The feature tree can represent the commonality and variability among reusable software artifacts~\cite{hariri2013supporting}. It can organize the artifacts in a hierarchical structure based on their functionality~\cite{reiser2007multi}. There have been various studies on extracting requirements features~\cite{guzman2014users}~\cite{ferrari2013mining}~\cite{kumaki2012supporting}~\cite{acher2012extracting}. Guzman et al.~\cite{guzman2014users} used part-of-speech tagging to extract functional features from app store reviews, helping developers analyze user feedback and identify high-frequency features. Ferrari et al.~\cite{ferrari2013mining} proposed a method based on natural language processing and comparative analysis to automatically extract commonalities and variabilities from documents of competing products. Kumaki et al.~\cite{kumaki2012supporting} proposed a technique based on the vector space model to automatically analyze the commonalities and variabilities of the requirements and structural models of legacy software assets. Mathieu et al.~\cite{acher2012extracting} used the domain-specific language VarCell to extract feature models from tabular requirements descriptions, ensuring that the generated models accurately reflect the commonalities and variabilities between artifacts. However, these existing works~\cite{ferrari2013mining}~\cite{weston2009framework} primarily focus on extracting features from requirement documents for forward reuse. They have limited attention to reverse extraction from reusable artifact descriptions. In addition, they typically rely on traditional natural language processing techniques, \eg part-of-speech tagging and term weighting. These methods may lack sufficient performance. Thus, this work aims to leverage the powerful ability of LLMs in requirements understanding to extract features from reusable artifacts for reverse engineering.

\subsection{LLMs for Requirements Understanding}
Researchers have used LLMs to improve or automate various requirements-related activities~\cite{khan2025large}, including requirements elicitation, analysis, specification, and validation. For example, Gorer et al. utilized LLMs and prompt engineering to generate requirements interview scripts automatically~\cite{gorer2023generating}. Ren et al. combined a few-shot learning to leverage LLMs to understand user reviews and classify them into requirements and features~\cite{ren2024combining}. Camara et al use ChatGPT to understand requirements descriptions to generate UML models~\cite{camara2023assessment}. Jin et al.~\cite{jin2024mare} proposed a multi-agent collaboration framework to generate software requirements specifications from a rough idea. Jin et al.~\cite{jin2024chatmodeler} proposed a human and LLMs collaboration approach to perform requirements elicitation, specification, and validation. These works demonstrate the powerful ability of LLMs to understand requirements. Thus, it is also an interesting topic to explore the use of LLMs for constructing requirements feature trees. 
\section{Approach}
In this section, we present an LLM-based multi-level feature tree construction framework, named {\sc FTBuilder}. We formally define the overview of our {\sc FTBuilder} and describe the details in the following sections, including three modules and recursive construction.

\subsection{Overview}
\begin{figure*}
    \centering
    \includegraphics[width=0.9\linewidth]{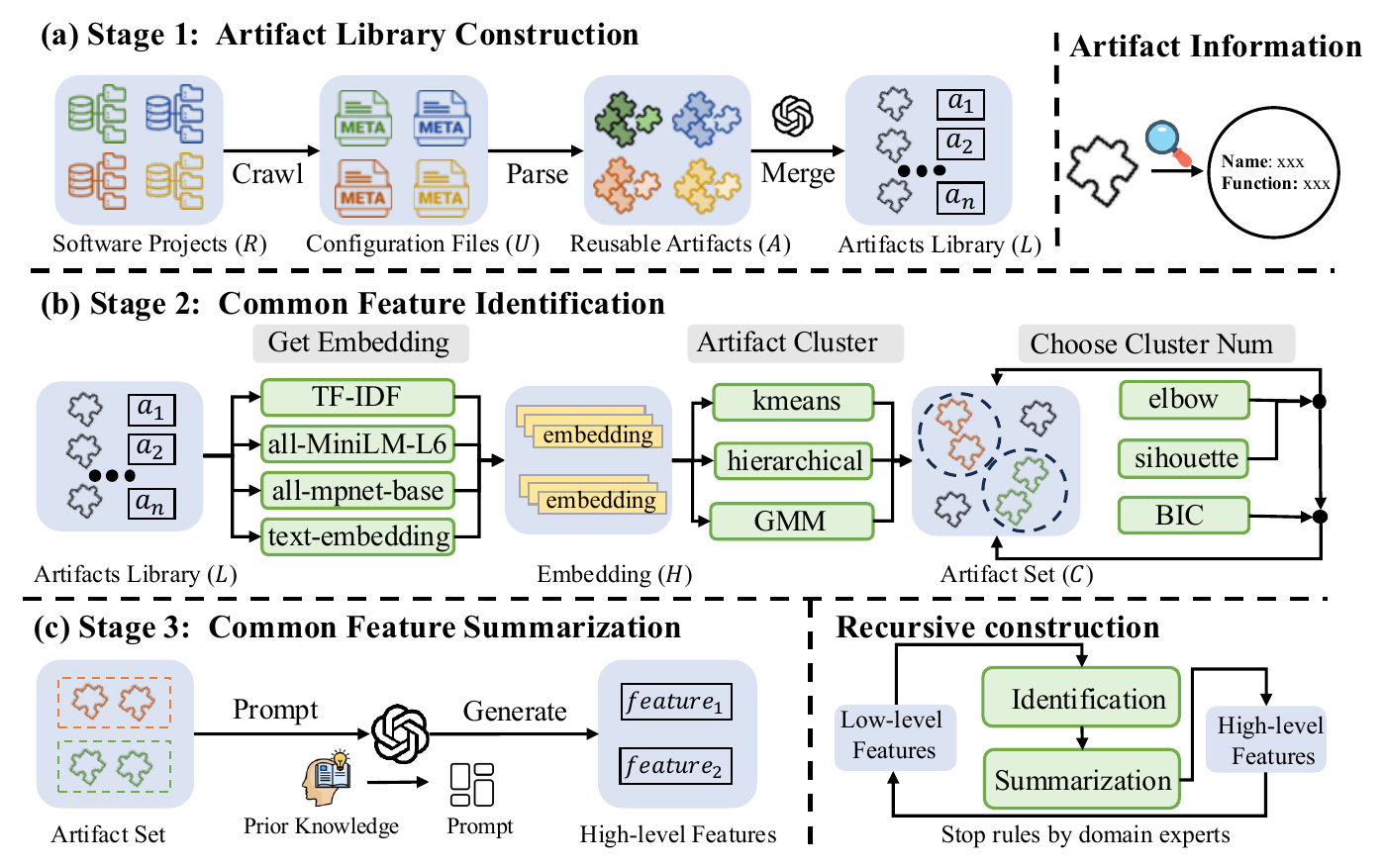}
    \caption{Overview of our {\sc FTBuilder}}
    \label{fig:approach}
\end{figure*}

Our approach aims to automatically construct a multi-level feature tree to manage reusable software artifacts based on domain-specific software projects. To achieve this, we decompose this task into three stages, including library construction, feature identification, and feature summarization. The three stages work in a pipeline as shown in Figure~\ref{fig:approach}.
\begin{itemize}
    \item \textbf{Library Construction.} Given domain-specific project repositories $R$, reusable artifacts $A$ are crawled and merged into a structured artifact library $L$. 
    \item \textbf{Feature Identification.} Based on the artifact library $L$, cluster algorithms are used to identify an artifact set $S$ that contains common features. 
    \item \textbf{Feature Summarization.} LLMs receive the functional descriptions of each artifact in $S$ and summarize their common features $F$.
\end{itemize}

\subsection{Structured Component Library} \label{sec:approach_library}
As shown in Figure~\ref{fig:approach}(a), this stage aims to construct a structured artifact library from open-source ecosystems or domain-specific software projects. Inspired by previous studies~\cite{jin2024first}, software projects typically use \textit{meta-packages}~\cite{web:metapackage} to define their reusable software artifacts. The details of these artifacts are stored in a specific file (\eg \textit{repomd.xml} for Linux.). Therefore, we crawl this file (\ie configuration file) in a software project and parse the artifact information.

Specifically, we review domain-specific open-source repositories $R = \{r_1,r_2,...,r_l\}$ to identify the URLs of the configuration files $U = \{u_1,u_2,...,u_m\}$. We then employ the \textit{requests} library~\cite{web:requests} to scrape these configuration files from each URL and parse them to extract information about the artifacts $A = \{a_1,a_2,...,a_n\}$. Each artifact $a_i$ contains two key attributes, \ie name and functional description. 


Since the software project repositories $R$ may be developed by different teams, the same artifact may have different names and functional descriptions. Therefore, we should consolidate the crawled raw artifacts $A$. Specifically, we adopt a gradual expansion approach and employ GPT-4 to automate this process. Initially, we set the library $L$ to be empty. We then design a prompt $P_c$ and use GPT-4 to assess whether each artifact $a_i$ already exists in the library $L$. If it does, we skip the artifact; Otherwise, we add it to the library $L$.


\begin{Prompt}{Prompt $P_c$ for Artifact Library Construction}
Artifact Library T: {$L$} 

Artifact N: {$a_i$}

Please judge if artifact N exists in Artifact Library T. 

A. Exists. B Not Exist
\end{Prompt}


\subsection{Common Feature Identification}
As shown in Figure~\ref{fig:approach}(b), the goal of this stage is to identify artifacts that contain common features from the constructed artifact library. We consider this procedure as a cluster task, where the main purpose is to group artifacts $a_i$ based on similarities in their functional descriptions. We first represent the descriptions of artifacts as semantic embeddings $H = [h_1,h_2,...,h_n]$ using embedding techniques (ET), such as TF-IDF and pre-trained LLMs. These vectors allow us to measure the similarity between artifacts.

\begin{equation}
\begin{aligned}
    H &= ET([a_1,a_2,...,a_n]) 
\end{aligned}
\end{equation}

We then employ clustering algorithms to group artifacts. We feed the embedding vectors $H$ into a clustering algorithm (CA). It divides these embedding vectors $H$ into $k$ clusters $C = [C_1,C_2,...,C_k]$. Artifacts within the same cluster are considered to have a common feature. 

\begin{equation}
\begin{aligned}
    C &= CA([h_1,h_2,...,h_n]) 
\end{aligned}
\end{equation}

This stage involves various design choices of the ET, CA, and the selection of the cluster number (CN). These choices can affect the quality of constructed feature trees and the best solution is closely related to the domain-specific data. Thus, we develop 24 solutions and provide support to select the optimal approach automatically. The developed solutions cover four ETs, three CAs, and three CNs~\footnote{Some combinations of ET, CA, and CN techniques are incompatible.}. A detailed introduction to these choices can be found in Section~\ref{sec:choice}.

\subsection{Common Feature Summarization}
As shown in Figure~\ref{fig:approach}(c), the goal of this stage is to generate the high-level common feature for each cluster $C_i$ identified in the previous stage. To achieve this, we employ GPT-4 to understand the functional descriptions of artifacts and summarize their shared characteristics into high-level common features. Specifically, we construct a prompt template as shown in $P_s$ based on prior knowledge. Then we pass the functional descriptions of artifacts in each cluster $C_i$ into the prompt template. GPT-4 receives the constructed prompt $P_i$ and generates a high-level common feature $F_i$ across the artifacts in the cluster.


\begin{Prompt}{Prompt $P_s$ for Common Feature Summarization}
Based on the following sub-features, please generate a parent common feature that can cover these sub-features. 
The sub-features are: \{\textit{the descriptions of artifacts in $C_i$}\}
Please only output the common feature in the format of `feature name: feature description:'.

\end{Prompt}

\subsection{Recursive Construction}
{\sc FTBuilder} recursively applies the identification and summarization stages to construct a multi-level feature tree. The recursion process continues until specific stopping criteria, which can be defined based on domain-specific expertise. The criteria typically include the maximum depth of the feature tree and the maximum number of features at the highest level.


\section{Study Design}
To evaluate the performance of the {\sc FTBuilder} framework, we conduct a multi-aspect study to answer four research questions (RQs). In this section, we describe the details of our study, including research questions, alternative solutions, datasets, and metrics.

\subsection{Research Questions}

\textbf{RQ1: How do the alternative solutions in {\sc FTBuilder} perform in feature tree construction?} We conduct an empirical study to evaluate the 24 alternative solutions using the Linux software package ecosystem (Section~\ref{sec:dataset}). Their effectiveness is evaluated using the silhouette score~\cite{vrezankova2018different} and Gvalue score~\cite{jin2024first} (Section~\ref{sec:metrics}).

\textbf{RQ2: How does the quality of the feature tree constructed by the best solutions compare to manual construction?} We apply the best solution in RQ1 to the Linux software package ecosystem. Then, we evaluate the quality of the constructed feature tree and compare it with the official feature tree using the coefficient score and Gvalue score. 

\textbf{RQ3: Does the feature tree constructed by {\sc FTBuilder} reduce practitioners' time in selecting artifacts?} We create a dataset named {\sc ArtSel} to simulate artifact selection based on requirements using the Linux software package ecosystem (Section~\ref{sec:dataset}). Three Linux developers are invited to select packages for given requirements in {\sc ArtSel}. We calculate and compare the cost time of each developer within the official feature tree and the one constructed by {\sc FTBuilder}.

\textbf{RQ4: Does the feature tree constructed by {\sc FTBuilder} improve the precision of LLMs in artifacts recommendation?} LLMs can recommend artifacts based on given requirements. We use {\sc ARTSEL} to evaluate and compare the precision of GPT-4 in artifact recommendation using the official tree and the one constructed by {\sc FTBuilder}.

\subsection{Solutions}~\label{sec:choice}
We develop 24 alternative solutions under the {\sc FTBuilder} framework for multi-level feature tree construction. They cover 4 embedding techniques, 3 cluster algorithms, and 3 strategies for selecting cluster numbers. 

\textbf{Embedding Techniques (ET).} We employ three types of advanced techniques. \textit{(1) Statistical Models.} TF-IDF~\cite{ramos2003using} calculates the importance of words by considering their frequency in an artifact description and their inverse frequency across all artifact descriptions. \textit{(2) Traditional Pre-trained Language Models.} SentenceTransformer~\cite{reimers-2020-multilingual-sentence-bert} provides various pre-trained models for computing vectors. We select the two most downloaded models, \ie all-MiniLM-L6 and all-mpnet-base. \textit{(3) LLM Embedding Models.} We select OpenAI's advanced embedding model, \ie text-embedding-ada-002~\cite{web:text-embedding}.

\textbf{Cluster Algorithms (CA).} We select three cluster algorithms: K-means~\cite{hartigan1979algorithm}, Gaussian Mixture Models (GMM)~\cite{reynolds2009gaussian}, and Hierarchical Clustering~\cite{murtagh2011methods}. K-means divides data into k clusters by minimizing the sum of squared distances between data points and their corresponding cluster centers. It requires the number of clusters $k$ to be specified beforehand. GMM is a probabilistic method that models data as a mixture of multiple gaussian distributions. Each cluster is represented by both its center and its shape, which allows GMM to handle complex clusters. However, the number of clusters $k$ still needs to be specified in advance. Hierarchical Clustering builds a tree structure by merging pairs of data points from the bottom up. It does not need to set the number of clusters in advance. 

\textbf{Select Cluster Numbers.} To determine the optimal number of clusters, we used three common methods: the Elbow Method, the Silhouette Method, and the Bayesian Information Criterion (BIC). The elbow method plots the number of clusters against the sum of squared errors (SSE). As the number of clusters increases, SSE decreases but eventually plateaus. The point where the reduction slows significantly is typically considered the optimal number of clusters. The silhouette method measures how similar a data point is to others in the same cluster compared to those in other clusters. A higher silhouette score indicates better clustering. The number of clusters that yield the highest silhouette score is selected.  BIC balances model fit and complexity. In clustering, a lower BIC value suggests a better model. We select the cluster number corresponding to the minimum BIC value.

\subsection{Dataset}~\label{sec:dataset}
We conduct experiments on an artifacts reuse dataset named {\sc ArtSel} created by this work. The {\sc ArtSel} uses the reusable \textit{group} artifacts from the open-source Linux distributions. 

\textbf{Artifacts Collection.}~\label{sec:collectdata}
Inspired by previous work~\cite{jin2024first}, this paper selected the same five widely used Linux distributions, including Fedora, CentOS, OpenEuler, Anolis, and OpenCloudOS. Reusable \textit{group} artifacts are collected from the official or widely available mirrors of these distributions. Each \textit{group} artifact represents a functional component to fulfill a specific requirement. Then, we parse the information of each \textit{group artifact}, including its name and functional description. Table~\ref{tab:data} presents the selected versions of Linux distributions and the statistics of collected \textit{group} artifacts from each distribution. Subsequently, the consolidation process in Section~\ref{sec:approach_library} is applied to merge these artifacts and construct a structured reusable artifact library. In total, the library contains 237 reusable \textit{group} artifacts. 

\begin{table}[]
    \centering
    \caption{The Artifacts Collection Source}
    \begin{tabular}{lllll}
\toprule
\textbf{Linux}  & \textbf{Version} & \textbf{Company} & \textbf{Mirror} & \textbf{\#Group}\\ \midrule
Fedora      & 40      & Red Hat           & Fedora & 158     \\
CentOS      & 7       & Red Hat           & CentOS   & 88      \\
OpenEuler   & 23.09   & Hua Wei           & Tsinghua & 52      \\
Anolis      & 8.9     & OpenAnolis        & Aliyun   & 74      \\
OpenCloudOS & 9.2     & China Electronics & CloudOS & 42      \\ \bottomrule
\end{tabular}
    \label{tab:data}
\end{table}

\textbf{Dataset Construction.}
The {\sc ARTSEL} dataset is designed to evaluate the effectiveness of selecting or recommending reusable artifacts based on specific requirements. Thus, each sample in the dataset should include a natural language requirement description $R$ and the corresponding reusable artifacts $A$ that satisfy the requirement. To construct the {\sc ARTSEL}, a group was randomly selected from the structured artifact library. The first author wrote a requirement description from a user perspective based on the group's functional description. The written requirements description should be able to be satisfied by the selected group artifact. This process was repeated 15 times, resulting in the {\sc ARTSEL} dataset containing 15 samples. To ensure the quality of the dataset, three Linux domain experts conduct a review process to verify the accuracy and relevance of the requirements to the artifacts. After three rounds of review and revision, all three experts endorsed each test sample in {\sc ARTSEL}.

\subsection{Metrics}~\label{sec:metrics}
We employ two metrics to evaluate the quality of constructed feature trees in RQ1 and RQ2.

\begin{itemize}
    \item \textbf{Silhouette Score (SS)} measures the similarity of a feature to other features under the same parent and its distinction from features under different parents~\cite{vrezankova2018different}. A higher silhouette score indicates a more coherent and well-structured feature tree. Specifically, the silhouette score is calculated as follows.

    \begin{equation}
        s(f_i) = \frac{b(f_i)- a(f_i)}{max(a(f_i),b(f_i))} 
    \end{equation}

    \begin{equation}
        S = \frac{1}{N} \sum_{i=1}^{n} s(f_i)
    \end{equation}

    where $f_i$ represents the $i$-th feature, $a(f_i)$ is the average distance between feature $f_i$ and other features under the same parent, $b(f_i)$ is the average distance between feature $f_i$ and the features in the closest parent, $s(f_i)$ is the silhouette score for feature $f_i$.
    \item \textbf{Gvalue Score (GS)} is a comprehensive metric that can be used to evaluate the feature tree. It can measure the rationality of the feature tree structure and reflect whether the parent feature covers the child features. A higher value indicates a higher-quality feature tree. The calculation method can be found in its paper~\cite{jin2024first}.
\end{itemize}

To evaluate the effectiveness of the feature tree in improving practitioners' efficiency in RQ3, we use the \textbf{Average Time Cost} of selecting the correct artifacts as the evaluation metric. Specifically, practitioners are provided with a requirement description and an artifact library. They are invited to select artifacts that satisfy the given requirement. We record and calculate their average time cost. In addition, we use \textbf{Precision} to evaluate the improvement of the feature tree on the artifacts recommendation task in RQ4. Specifically, we compare the recommended artifacts with the correct artifacts and calculate the proportion of correctly recommended artifacts out of the total recommended artifacts.

\section{Results and Analysis}
\textbf{RQ1: How do the alternative solutions in {\sc FTBuilder} perform in feature tree construction?} 

\textbf{Setup.} The 24 alternative solutions (Section~\ref{sec:choice}) are used to construct the multi-level feature tree for the collected \textit{group} artifacts (Section~\ref{sec:collectdata}). The stopping criterion for recursive construction is that the number of features at the highest level should not be less than 4~\cite{tan2023case}. We evaluate the quality of the constructed feature trees. The evaluation metrics are described in Section~\ref{sec:metrics}, \ie the silhouette score and Gvalue score. For all metrics, higher scores represent better performance.  

\textbf{Results.} Table~\ref{tab:empirical} shows the experimental results of the 24 alternative solutions, including the statistics and evaluation of their constructed feature trees. ``\#L'' and ``\#N'' denote the feature tree's number of layers and nodes, respectively. We can find that the best solution is to use text-embedding-ada-002 to obtain embedding vectors, GMM for clustering, and BIC to select the number of clusters. The best solution achieves a silhouette score of 0.067 and a GValue score of 0.56. 

\begin{table}[]
    \centering
    \setlength{\tabcolsep}{5pt}
    \caption{Empirical Study on 24 alternative solutions in {\sc FTBuilder}}
    \begin{tabular}{ccccccc}
\toprule
\multicolumn{3}{c}{\textbf{Solutions}}                 & \multicolumn{2}{c}{\textbf{Tree}} & \multicolumn{2}{l}{\textbf{Metrics}} \\ \midrule
\textbf{ET} & \textbf{CA} & \textbf{CN}  & \textbf{\#L} & \textbf{\#N}  & \textbf{SS}            & \textbf{GS}          \\ \midrule
TF-IDF             & kmeans       & elbow     & 3           & 255        & 0.027         & 0.45        \\
TF-IDF             & kmeans       & sihouette & 3           & 264        & 0.022         & 0.51        \\
TF-IDF             & GMM          & elbow     & 3           & 244        & -0.002        & 0.46        \\
TF-IDF             & GMM          & sihouette & 3           & 268        & 0.013         & 0.54        \\
TF-IDF             & GMM          & BIC       & 2           & 256        & 0.015         & 0.46        \\
TF-IDF             & hierarchical & -         & 3           & 308        & 0.013         & 0.47        \\ \midrule
all-MiniLM-L6      & kmeans       & elbow     & 2           & 246        & 0.043         & 0.46        \\
all-MiniLM-L6      & kmeans       & sihouette & 2           & 250        & 0.057         & 0.48        \\
all-MiniLM-L6      & GMM          & elbow     & 2           & 243        & 0.028         & 0.42        \\
all-MiniLM-L6      & GMM          & sihouette & 2           & 242        & 0.038         & 0.43        \\
all-MiniLM-L6      & GMM          & BIC       & 2           & 246        & 0.037         & 0.44        \\
all-MiniLM-L6      & hierarchical & -         & 3           & 309        & 0.001         & 0.73        \\ \midrule
all-mpnet-base     & kmeans       & elbow     & 2           & 247        & 0.033         & 0.45        \\
all-mpnet-base     & kmeans       & sihouette & 3           & 264        & 0.044         & 0.55        \\
all-mpnet-base     & GMM          & elbow     & 3           & 246        & 0.081         & 0.46        \\
all-mpnet-base     & GMM          & sihouette & 3           & 258        & 0.047         & 0.48        \\
all-mpnet-base     & GMM          & BIC       & 3           & 252        & 0.042         & 0.48        \\
all-mpnet-base     & hierarchical & -         & 3           & 308        & 0.012         & 0.46         \\ \midrule
text-embedding & kmeans       & elbow     & 2           & 247        & 0.066         & 0.47        \\
text-embedding & kmeans       & sihouette & 2           & 261        & 0.053         & 0.52        \\
text-embedding & GMM          & elbow     & 3           & 247        & 0.047         & 0.47        \\
text-embedding & GMM          & sihouette & 3           & 264        & 0.061         & 0.50        \\
{\color[HTML]{FE0000}\textbf{text-embedding}} & {\color[HTML]{FE0000}\textbf{GMM}}          & {\color[HTML]{FE0000}\textbf{BIC}}       & {\color[HTML]{FE0000}\textbf{3}}           & {\color[HTML]{FE0000}\textbf{245}}        &{\color[HTML]{FE0000}\textbf{0.067}}         & {\color[HTML]{FE0000}\textbf{0.56}}        \\
text-embedding & hierarchical & -         & 4            & 390           & 0.023           & 0.47            \\ \bottomrule
\end{tabular}
    \label{tab:empirical}
\end{table}

\textbf{RQ2: How does the quality of the feature tree from the best solution compare to manual construction?} 

\textbf{Setup.} We evaluate and compare the manual feature tree and our constructed feature tree with the optimal solution for the Linux distribution ecosystem. The evaluation metrics are also the silhouette score and Gvalue score. 

\begin{figure*}
    \centering
    \includegraphics[width=0.99\linewidth]{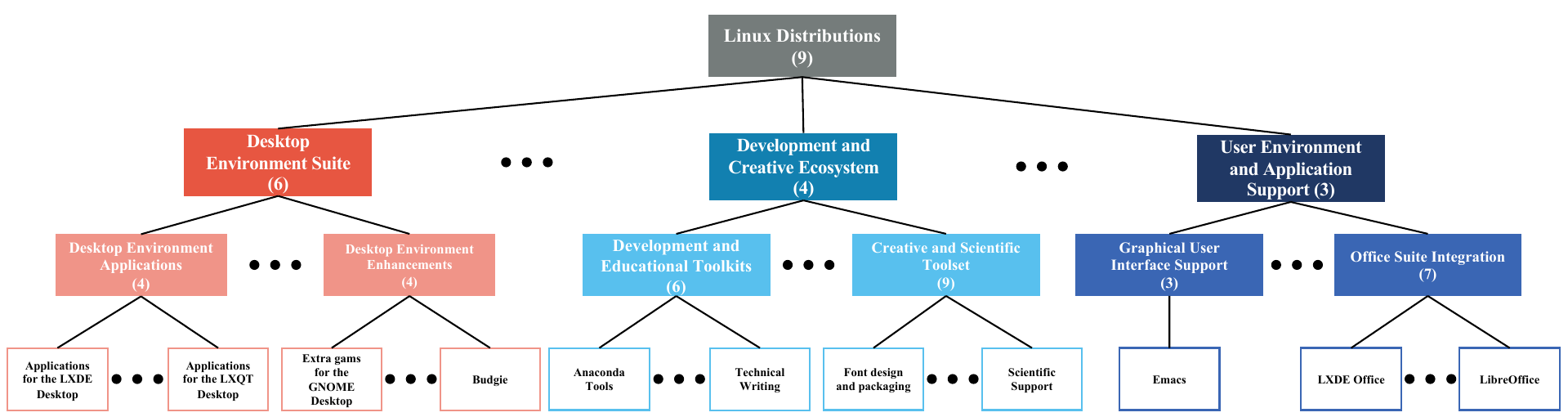}
    \caption{A fragment of the constructed multi-level feature tree for Linux distributions.}
    \label{fig:feature_tree}
\end{figure*}

\textbf{Results.} Table~\ref{tab:tree quality} demonstrates the comparative results about the quality of trees, and Figure~\ref{fig:feature_tree} shows a fragment of the constructed multi-level feature tree for Linux distributions. 
Detailed information and visualizations of the two trees can be found in our replication package~\cite{web:code}. We can observe that our feature tree outperforms the manually constructed official feature tree, showing a 9\% improvement in the silhouette score and an 11\% increase in the Gvalue score. This improvement is due to a more compact and efficient structure, with fewer layers and nodes in our feature tree. This suggests that our approach reduces redundancy and improves the organization and relationships between features.

\begin{table}[]
    \centering
    \caption{Comparison of the quality of the official feature tree and constructed feature tree}
    \begin{tabular}{lcccc}
\toprule
\textbf{Feature Tree}                       & \multicolumn{1}{l}{\textbf{Layer}} & \multicolumn{1}{l}{\textbf{Node}} & \multicolumn{1}{l}{\textbf{SS}} & \multicolumn{1}{l}{\textbf{GS}}  \\ \midrule
Official                           &   5                        &  723                        & 0.059                         & 0.50                        \\
Ours                               & 3                         & 245                      & 0.067                         & 0.56                        \\ \midrule
{\color[HTML]{FE0000} Improvement Ratio} & {\color[HTML]{FE0000} -}  & {\color[HTML]{FE0000} -} & {\color[HTML]{FE0000} 9\%}    & {\color[HTML]{FE0000} 11\%} \\ \bottomrule
\end{tabular}
    \label{tab:tree quality}
\end{table}

\textbf{RQ3: Does the feature tree constructed by {\sc FTBuilder} reduce practitioners' time in selecting reusable artifacts?}

\textbf{Setup.} We conduct interviews with three Linux practitioners (\ie A, B, and C). They are all computer science Ph.D. students and are not co-authors. During the interviews, each participant is provided with the natural language requirements and an artifact library. They are asked to select the required artifacts with the guidance of the constructed feature tree and official tree, separately. The average time spent by each practitioner on each test sample is recorded and compared. 

\begin{table}[]
    \centering
    \caption{The cost time on selecting artifacts for practitioners}
    \begin{tabular}{ccccc}
\toprule
\multicolumn{1}{c}{\multirow{2}{*}{\textbf{Approach}}} & \multicolumn{4}{c}{\textbf{Cost Time (min/sample)}}                                                                       \\ \cmidrule{2-5} 
\multicolumn{1}{c}{}                          & \multicolumn{1}{c}{\textbf{A}} & \multicolumn{1}{c}{\textbf{B}} & \multicolumn{1}{c}{\textbf{C}} & \multicolumn{1}{c}{\textbf{Average}} \\ \midrule
With Official Tree                            &    6                  &         4              &         5              &          5                  \\
With Ours Tree                                &     4                  &         3              &          4             &           3.67                  \\ \midrule
{\color[HTML]{FE0000} Improvement Ratio}                                   &     {\color[HTML]{FE0000} 33\%}                  &        {\color[HTML]{FE0000} 25\%}               &         {\color[HTML]{FE0000} 20\%}              &           {\color[HTML]{FE0000} 26\%}                  \\ \bottomrule
\end{tabular}
    \label{tab:time_cost}
\end{table}

\textbf{Results.} Table~\ref{tab:time_cost} presents the average cost time on select artifacts for each practitioner with the official and constructed feature tree. We can observe that the feature tree constructed by {\sc FTBuilder} consistently reduces the selection time across all practitioners compared to the official tree. Specifically, the average time reduction across all practitioners is about 26\%. This suggests that the constructed feature tree effectively helps practitioners select artifacts more efficiently. 

\textbf{RQ4: Does the feature tree constructed by {\sc FTBuilder} improve the performance of LLMs in artifact recommendation?} 

\textbf{Setup.} We use the {\sc ARTSEL} dataset to evaluate and compare the efficiency and accuracy of LLMs (\ie GPT-4 and Deepseek) on the reusable artifact recommend task with the guidance of the official tree and the constructed tree. The metrics are the time cost and accuracy (Section~\ref{sec:metrics}). 

\textbf{Results.} Table~\ref{tab:llm} represents the comparative results for the artifact recommendation using LLMs. We can observe that the feature tree constructed by {\sc FTBuilder} significantly improves the performance of LLMs in artifact recommendation tasks. The constructed tree leads to faster recommendation time and higher accuracy in both LLMs. Specifically, the constructed tree reduces time by 48\% and improves accuracy by 235\% for GPT-4o, while it reduces time by 13\% and improves accuracy by 237\% for DeepSeek-R1.
\begin{table}[]
    \centering
    \caption{Time and Accuracy on Artifact recommendation by LLMs}
    \begin{tabular}{cccc}
\toprule
\textbf{LLM}                       & \textbf{Approach}            & \textbf{Time(min)} & \textbf{Accuracy} \\ \midrule
\multirow{3}{*}{GPT-4o}    & with official tree & 9.28     &  20\%       \\
                          & with our tree      &  4.85    &  67\%        \\ \cmidrule{2-4} 
                          & {\color[HTML]{FE0000} Improvement Ratio}        & {\color[HTML]{FE0000} 48\%}     & {\color[HTML]{FE0000} 235\%}         \\ \midrule
\multirow{3}{*}{DeepSeek-R1}    & with official tree &  71.34   &  13\%       \\
                          & with our tree      & 61.72 &  45\%        \\ \cmidrule{2-4} 
                          & {\color[HTML]{FE0000} Improvement Ratio}        &  {\color[HTML]{FE0000} 13\%}      & {\color[HTML]{FE0000} 237\%}         \\ \bottomrule
\end{tabular}
    \label{tab:llm}
\end{table}



\section{Research Plan}
In this section, we outline our plans for continuing our research. Our efforts will focus on a more extensive evaluation of our existing work, including involving additional software ecosystems, constructing a large-scale artifact reuse dataset, and evaluating more LLMs. The detailed research plans are described below.

\textbf{Involving more software ecosystems or domain-specific projects.} Our current research is focused on the Linux software package ecosystem. Thus, we plan to extend our {\sc FTBuilder} to cover other open-source ecosystems (\eg JavaScript) and domain-specific projects (\eg aerospace). This expansion will include different types of artifacts and different application domains. By extending to these ecosystems and domains, we aim to validate the versatility of our {\sc FTBuilder} and assess its performance in diverse real-world situations.

\textbf{Constructing a large-scale artifact reuse and recommendation dataset.} The current {\sc ARTSEL} dataset has a limited number of test samples and includes only one type of artifact (\ie \textit{group}). To enhance the utility and coverage, we plan to expand this dataset by collecting and creating more requirements-artifacts pairs from various ecosystems. This will involve increasing the diversity of artifact types (\eg packages and code snippets) and increasing the number of samples in the dataset. the scale of the dataset. By extending this dataset, we can provide a more comprehensive evaluation for our {\sc FTBuilder} and support future development and evaluation of artifact reuse and recommendation techniques. 

\textbf{Conducting a more comprehensive evaluation.} For the evaluation of manual artifact selection efficiency (RQ3), We plan to extend our current results by conducting experiments on the new large-scale dataset and inviting more practitioners from diverse backgrounds to participate in the validation process. For artifact recommendation using LLMs (RQ4), we plan to broaden our evaluation to include a wider range of advanced LLMs (\eg LLama4, Qwen2.5, Claude). We aim to compare their performance in artifact recommendation tasks and analyze the impact of constructed feature trees.

\textbf{Practice in industrial scenarios.} We plan to collaborate with industrial partners to apply {\sc FTBuilder} in real-world projects and asses its effectiveness. We will design a questionnaire to collect the experiences and feedback from practitioners. This can allow us to validate the practical impact in a real-world setting and conduct case studies to analyze directions for further improvement.

\section{conclusion}
In this paper, we propose an automatic LLM-based multi-level feature tree construction framework named {\sc FTBuilder} for domain-specific reusable artifacts management. The framework consists of three stages: Library Construction, Feature Identification, and Feature Summarization. It recursively applies the identification and summarization stages to construct a multi-level feature tree from the bottom up. We have developed 24 alternative solutions under the {\sc FTBuilder} and made them available to support practitioners for their respective projects. To validate the effectiveness, we create a small-scale artifact reuse and recommendation dataset named {\sc ARTSEL} and conduct experiments from multiple aspects to evaluate it. The results show that the constructed tree by our {\sc FTBuilder} outperforms the official feature tree. Moreover, it can reduce the practitioners' time in selecting artifacts and improve the precision of LLMs in artifact recommendation.

\section{ACKNOWLEDGMENT}
This research is supported by the National Natural Science Foundation of China (62192731, 62192730).



\useunder{\uline}{\ul}{}

\normalem
\balance
\bibliographystyle{IEEEtran}
\bibliography{main}

\begin{thebibliography}{10}
\providecommand{\url}[1]{#1}
\csname url@samestyle\endcsname
\providecommand{\newblock}{\relax}
\providecommand{\bibinfo}[2]{#2}
\providecommand{\BIBentrySTDinterwordspacing}{\spaceskip=0pt\relax}
\providecommand{\BIBentryALTinterwordstretchfactor}{4}
\providecommand{\BIBentryALTinterwordspacing}{\spaceskip=\fontdimen2\font plus
\BIBentryALTinterwordstretchfactor\fontdimen3\font minus \fontdimen4\font\relax}
\providecommand{\BIBforeignlanguage}[2]{{%
\expandafter\ifx\csname l@#1\endcsname\relax
\typeout{** WARNING: IEEEtran.bst: No hyphenation pattern has been}%
\typeout{** loaded for the language `#1'. Using the pattern for}%
\typeout{** the default language instead.}%
\else
\language=\csname l@#1\endcsname
\fi
#2}}
\providecommand{\BIBdecl}{\relax}
\BIBdecl

\bibitem{frakes2005software}
W.~B. Frakes and K.~Kang, ``Software reuse research: Status and future,'' \emph{IEEE transactions on Software Engineering}, vol.~31, no.~7, pp. 529--536, 2005.

\bibitem{lim1994effects}
W.~C. Lim, ``Effects of reuse on quality, productivity, and economics,'' \emph{IEEE software}, vol.~11, no.~5, pp. 23--30, 1994.

\bibitem{mohagheghi2004empirical}
P.~Mohagheghi, R.~Conradi, O.~M. Killi, and H.~Schwarz, ``An empirical study of software reuse vs. defect-density and stability,'' in \emph{Proceedings. 26th International Conference on Software Engineering}, 2004, pp. 282--291.

\bibitem{mohagheghi2007quality}
P.~Mohagheghi and R.~Conradi, ``Quality, productivity and economic benefits of software reuse: a review of industrial studies,'' \emph{Empirical Software Engineering}, vol.~12, pp. 471--516, 2007.

\bibitem{wittern2016look}
E.~Wittern, P.~Suter, and S.~Rajagopalan, ``A look at the dynamics of the javascript package ecosystem,'' in \emph{Proceedings of the 13th international conference on mining software repositories}, 2016, pp. 351--361.

\bibitem{web:npm}
NPM, ``Node package manager for javascript,'' \url{https://www.npmjs.com/}, 2025.

\bibitem{mockus2000case}
A.~Mockus, R.~T. Fielding, and J.~Herbsleb, ``A case study of open source software development: the apache server,'' in \emph{Proceedings of the 22nd international conference on Software engineering}, 2000, pp. 263--272.

\bibitem{jin2024first}
D.~Jin, N.~Li, K.~Yang, M.~Zhou, and Z.~Jin, ``A first look at package-to-group mechanism: An empirical study of the linux distributions,'' \emph{arXiv preprint arXiv:2410.10131}, 2024.

\bibitem{bakar2015feature}
N.~H. Bakar, Z.~M. Kasirun, and N.~Salleh, ``Feature extraction approaches from natural language requirements for reuse in software product lines: A systematic literature review,'' \emph{Journal of Systems and Software}, vol. 106, pp. 132--149, 2015.

\bibitem{reiser2007multi}
M.-O. Reiser and M.~Weber, ``Multi-level feature trees: A pragmatic approach to managing highly complex product families,'' \emph{Requirements Engineering}, vol.~12, pp. 57--75, 2007.

\bibitem{hariri2013supporting}
N.~Hariri, C.~Castro-Herrera, M.~Mirakhorli, J.~Cleland-Huang, and B.~Mobasher, ``Supporting domain analysis through mining and recommending features from online product listings,'' \emph{IEEE Transactions on Software Engineering}, vol.~39, no.~12, pp. 1736--1752, 2013.

\bibitem{guzman2014users}
E.~Guzman and W.~Maalej, ``How do users like this feature? a fine grained sentiment analysis of app reviews,'' in \emph{2014 IEEE 22nd international requirements engineering conference (RE)}, 2014, pp. 153--162.

\bibitem{ferrari2013mining}
A.~Ferrari, G.~O. Spagnolo, and F.~Dell'Orletta, ``Mining commonalities and variabilities from natural language documents,'' in \emph{Proceedings of the 17th International Software Product Line Conference}, 2013, pp. 116--120.

\bibitem{weston2009framework}
N.~Weston, R.~Chitchyan, and A.~Rashid, ``A framework for constructing semantically composable feature models from natural language requirements,'' in \emph{Proceedings of the 13th International Software Product Line Conference}, 2009, pp. 211--220.

\bibitem{davril2013feature}
J.-M. Davril, E.~Delfosse, N.~Hariri, M.~Acher, J.~Cleland-Huang, and P.~Heymans, ``Feature model extraction from large collections of informal product descriptions,'' in \emph{proceedings of the 2013 9th joint meeting on foundations of software engineering}, 2013, pp. 290--300.

\bibitem{kumaki2012supporting}
K.~Kumaki, R.~Tsuchiya, H.~Washizaki, and Y.~Fukazawa, ``Supporting commonality and variability analysis of requirements and structural models,'' in \emph{Proceedings of the 16th International Software Product Line Conference-Volume 2}, 2012, pp. 115--118.

\bibitem{yu2013mining}
Y.~Yu, H.~Wang, G.~Yin, and B.~Liu, ``Mining and recommending software features across multiple web repositories,'' in \emph{Proceedings of the 5th Asia-Pacific Symposium on Internetware}, 2013, pp. 1--9.

\bibitem{acher2012extracting}
M.~Acher, A.~Cleve, G.~Perrouin, P.~Heymans, C.~Vanbeneden, P.~Collet, and P.~Lahire, ``On extracting feature models from product descriptions,'' in \emph{Proceedings of the 6th International Workshop on Variability Modeling of Software-Intensive Systems}, 2012, pp. 45--54.

\bibitem{kojima2022large}
T.~Kojima, S.~S. Gu, M.~Reid, Y.~Matsuo, and Y.~Iwasawa, ``Large language models are zero-shot reasoners,'' \emph{Advances in neural information processing systems}, vol.~35, pp. 22\,199--22\,213, 2022.

\bibitem{gorer2023generating}
B.~G{\"o}rer and F.~B. Aydemir, ``Generating requirements elicitation interview scripts with large language models,'' in \emph{IEEE 31st International Requirements Engineering Conference Workshops}, 2023, pp. 44--51.

\bibitem{jin2024evaluation}
D.~Jin, S.~Zhao, Z.~Jin, X.~Chen, C.~Wang, Z.~Fang, and H.~Xiao, ``An evaluation of requirements modeling for cyber-physical systems via llms,'' \emph{arXiv preprint arXiv:2408.02450}, 2024.

\bibitem{camara2023assessment}
J.~C{\'a}mara, J.~Troya, L.~Burgue{\~n}o, and A.~Vallecillo, ``On the assessment of generative ai in modeling tasks: an experience report with chatgpt and uml,'' \emph{Software and Systems Modeling}, vol.~22, no.~3, pp. 781--793, 2023.

\bibitem{lubos2024leveraging}
S.~Lubos, A.~Felfernig, T.~N.~T. Tran, D.~Garber, M.~El~Mansi, S.~P. Erdeniz, and V.-M. Le, ``Leveraging llms for the quality assurance of software requirements,'' in \emph{32nd International Requirements Engineering Conference}, 2024, pp. 389--397.

\bibitem{lutze2024generating}
R.~Lutze and K.~Waldh{\"o}r, ``Generating specifications from requirements documents for smart devices using large language models (llms),'' in \emph{International Conference on Human-Computer Interaction}, 2024, pp. 94--108.

\bibitem{vrezankova2018different}
H.~{\v{R}}ezankov{\'a}, ``Different approaches to the silhouette coefficient calculation in cluster evaluation,'' in \emph{21st international scientific conference AMSE applications of mathematics and statistics in economics}, 2018, pp. 1--10.

\bibitem{web:rpmfinder}
RPM, ``https://rpmfind.net/linux/rpm/groups.html,'' \url{https://rpmfind.net/linux/RPM/Groups.html}, 2025.

\bibitem{web:code}
``Our code and constructed trees,'' \url{https://github.com/jdm4pku/FTBuilder}.

\bibitem{mili1995reusing}
H.~Mili, F.~Mili, and A.~Mili, ``Reusing software: Issues and research directions,'' \emph{IEEE transactions on Software Engineering}, vol.~21, no.~6, pp. 528--562, 1995.

\bibitem{gill2006importance}
N.~S. Gill, ``Importance of software component characterization for better software reusability,'' \emph{ACM SIGSOFT Software Engineering Notes}, vol.~31, no.~1, pp. 1--3, 2006.

\bibitem{isakowitz1996supporting}
T.~Isakowitz and R.~J. Kauffman, ``Supporting search for reusable software objects,'' \emph{IEEE Transactions on Software engineering}, vol.~22, no.~6, pp. 407--423, 1996.

\bibitem{tomer2004evaluating}
A.~Tomer, L.~Goldin, T.~Kuflik, E.~Kimchi, and S.~R. Schach, ``Evaluating software reuse alternatives: a model and its application to an industrial case study,'' \emph{IEEE Transactions on Software Engineering}, vol.~30, no.~9, pp. 601--612, 2004.

\bibitem{rothenberger2003strategies}
M.~A. Rothenberger, K.~J. Dooley, U.~R. Kulkarni, and N.~Nada, ``Strategies for software reuse: A principal component analysis of reuse practices,'' \emph{IEEE Transactions on Software Engineering}, vol.~29, no.~9, pp. 825--837, 2003.

\bibitem{lee1997empirical}
N.-Y. Lee and C.~R. Litecky, ``An empirical study of software reuse with special attention to ada,'' \emph{IEEE Transactions on Software Engineering}, vol.~23, no.~9, pp. 537--549, 1997.

\bibitem{kazman1996scenario}
R.~Kazman, G.~Abowd, L.~Bass, and P.~Clements, ``Scenario-based analysis of software architecture,'' \emph{IEEE software}, vol.~13, no.~6, pp. 47--55, 1996.

\bibitem{khan2025large}
J.~A. Khan, S.~Qayyum, and H.~S. Dar, ``Large language model for requirements engineering: A systematic literature review,'' 2025.

\bibitem{ren2024combining}
S.~Ren, H.~Nakagawa, and T.~Tsuchiya, ``Combining prompts with examples to enhance llm-based requirement elicitation,'' in \emph{2024 IEEE 48th Annual Computers, Software, and Applications Conference}, 2024, pp. 1376--1381.

\bibitem{jin2024mare}
D.~Jin, Z.~Jin, X.~Chen, and C.~Wang, ``Mare: Multi-agents collaboration framework for requirements engineering,'' \emph{arXiv preprint arXiv:2405.03256}, 2024.

\bibitem{jin2024chatmodeler}
------, ``Chatmodeler: a human-machine collaborative and iterative requirements elicitation and modeling approach via large language models,'' \emph{J Comput Res Develop}, vol.~61, no.~02, pp. 338--350, 2024.

\bibitem{web:metapackage}
``Metapackage in linux distributions,'' \url{https://help.ubuntu.com/community/MetaPackages}.

\bibitem{web:requests}
Requests, ``Requests library: Http for humans,'' \url{https://requests.readthedocs.io/en/latest/}, 2025.

\bibitem{ramos2003using}
J.~Ramos \emph{et~al.}, ``Using tf-idf to determine word relevance in document queries,'' in \emph{Proceedings of the first instructional conference on machine learning}, vol. 242, no.~1, 2003, pp. 29--48.

\bibitem{reimers-2020-multilingual-sentence-bert}
N.~Reimers and I.~Gurevych, ``Making monolingual sentence embeddings multilingual using knowledge distillation,'' in \emph{Proceedings of the 2020 Conference on Empirical Methods in Natural Language Processing}, 11 2020.

\bibitem{web:text-embedding}
OpenAI, ``Text-embedding-ada-002,'' \url{https://openai.com/index/new-and-improved-embedding-model/}, 2025.

\bibitem{hartigan1979algorithm}
J.~A. Hartigan and M.~A. Wong, ``Algorithm as 136: A k-means clustering algorithm,'' \emph{Journal of the royal statistical society. series c (applied statistics)}, vol.~28, no.~1, pp. 100--108, 1979.

\bibitem{reynolds2009gaussian}
D.~A. Reynolds \emph{et~al.}, ``Gaussian mixture models.'' \emph{Encyclopedia of biometrics}, vol. 741, no. 659-663, p.~3, 2009.

\bibitem{murtagh2011methods}
F.~Murtagh and P.~Contreras, ``Methods of hierarchical clustering,'' \emph{arXiv preprint arXiv:1105.0121}, 2011.

\bibitem{tan2023case}
J.~Tan, L.~Zhang, J.~Meng, H.~Xue, Z.~Liu, Z.~Ding, and Q.~Jing, ``A case study of an automatic package layering algorithm for linux distributions,'' in \emph{Proceedings of the 2023 4th International Conference on Computing, Networks and Internet of Things}, 2023, pp. 67--74.

\end{thebibliography}

\end{document}